\documentclass[runningheads]{llncs}
\usepackage{graphicx}

\usepackage{times}
\usepackage{latexsym}

\usepackage{amsmath}
\usepackage{multirow}
\usepackage{diagbox}

\usepackage{url}
\begin{document}
\title{Discover Your Social Identity from What You Tweet:\\ a Content Based Approach}
%
%
\author{Binxuan Huang \and
Kathleen M. Carley}
\authorrunning{B. Huang \and K. Carley}
%
\institute{Carnegie Mellon University,\\ Pittsburgh Pa 15213, USA\\
\email{binxuanh@cs.cmu.edu}\\
\email{kathleen.carley@cs.cmu.edu}}
\maketitle              
\begin{abstract}
An identity denotes the role an individual or a group plays in highly differentiated contemporary societies. In this paper, our goal is to classify Twitter users based on their role identities. We first collect a coarse-grained public figure dataset automatically, then manually label a more fine-grained identity dataset. We propose a hierarchical self-attention neural network for Twitter user role identity classification. Our experiments demonstrate that the proposed model significantly outperforms multiple baselines. We further propose a transfer learning scheme that improves our model's performance by a large margin. Such transfer learning also greatly reduces the need for a large amount of human labeled data.
\keywords{Social Identity \and Twitter \and User Profiling \and Text Mining \and Neural Network}

\end{abstract}

\section{introduction}
An identity is a characterization of the role an individual takes on. It is often described as the social context specific personality of an individual actor or a group of people \cite{ashforth1989social}. Identities can be things like jobs (e.g. ``lawyer'', ``teacher''), gender (man, woman), or a distinguishing characteristic (e.g. ``a shy boy'', ``a kind man''). People with different identities tend to exhibit different behaviors in the social space \cite{callero1985role}. In this paper, we use role identity to refer to the roles individuals or groups play in society.

Specifically on social media platforms, there are many different kinds of actors using social media, e.g., people, organizations, and bots. Each type of actor has different motivations, different resources at their disposal, and may be under different internal policies or constraints on when they can use social media, how they can represent themselves, and what they can communicate. If we want to understand who is controlling the conversation and whom is being impacted, it is important to know what types of actors are doing what. 

To date, for Twitter, most research has separated types of actors largely based on whether the accounts are verified by Twitter or not \cite{hentschel2014finding}, or whether they are bots or not \cite{chu2010tweeting}.
Previous study has shown that separating Twitter users into bots and non-bots provides better understanding of U.S. presidential election online discussion \cite{bessi2016social}. Bessi and Ferrara reveal that social bots distort the 2016 U.S. presidential election online discussion and about one-fifth of the entire conversation comes from bots. However, a variety of different types of actors may be verified - e.g., news agencies, entertainment or sports team, celebrities, and politicians. Similarly, bots can vary - e.g., news bots and non-news bots. If we could classify the role identities of actors on Twitter, we could gain an improved understanding of who was doing the influencing and who was being influenced \cite{cha2010measuring}. For example, knowing the social roles of bots would enable a more in-depth analysis of bot activities in the diffusion process of disinformation, eg. whether bots pretend to be news agencies to persuade regular users.

Understanding the sender's role is critical for doing research on, and developing technologies to stop, disinformation \cite{carley2018social,NAP25335}.
Research has shown that disinformation has a greater reach if it is spread by news agencies and celebrities \cite{babcock2019}. 
Disinformation is generally thought to be promoted by bots \cite{babcock2018beaten,uyheng2019characterizing}; however, most tools for identifying bots have relatively low accuracy when used in the wild \cite{beskow2018bot}.  News reporters, news agencies and celebrities often look like bots. Separating them out gives a better understanding of the role of bots in promoting disinformation.  Assessing the extent to which official sites are communicating in a way that effectively counters disinformation also required identification of the sender's role. Thus, role identification is foundational for disinformation research

In this paper, the primary goal is to classify Twitter users based on their role identities on social media. 
First, we introduce two datasets for Twitter user identity classification. One is automatically collected from Twitter aiming at identifying public figures on social media. Another is a human labeled dataset for more fine-grained Twitter user identity classification, which includes identities like government officials, news reporters, etc. Second, we present a hierarchical self-attention neural network for Twitter user identity classification. In our experiments, we show our method achieves excellent results when compared to many strong classification baselines. Last but not least, we propose a transfer learning scheme for fine-grained user identity classification which boosts our model's performance a lot.


\section{Related Work}
Sociologists have long been interested in the usage of identities across various social contexts \cite{tajfel1974social}. As summarized in \cite{stryker2000past}, three relatively distinct usages of \textit{identity} exist in the literature. Some use identity to refer to the culture of a people \cite{calhoun1994social2}. Some use it to refer to common identification with a social category \cite{tajfel1982social}. While others use identity to refer to the role a person plays in highly differentiated contemporary societies. In this paper, we use the third meaning.  Our goal for identity classification is to separate actors with different roles in online social media.

Identity is the way that individuals and collectives are distinguished in their relations with others \cite{jenkins2014social}. Certain difficulties still exist for categorizing people into different groups based on their identities. Recasens et al. argue that identity should be considered to be varying in granularity and a categorical understanding would limit us in a fixed scope \cite{recasens2011identity}. While much work could be done along this line, at this time we adopt a coarse-grained labeling procedure, that only looks at major identities in the social media space.

Twitter, a popular online news and social networking site, is also a site that affords interactive identity presentation to unknown audiences. As pointed out by Robinson et al., individuals form new cyber identities on the internet, which are not necessarily the way they would be perceived offline \cite{robinson2007cyberself}. A customized identity classifier is needed for online social media like Twitter.

A lot of research has tried to categorize Twitter users based on certain criteria \cite{rangel2015overview}, like gender \cite{burger2011discriminating}, location \cite{huang2019hierarchical,huang2017predicting,zhang2017rate}, occupation \cite{hu2016language,preoctiuc2015analysis}, and political orientation \cite{colleoni2014echo}. Another similar research topic is bot detection \cite{chu2010tweeting}, where the goal is to identify automated user accounts from normal Twitter accounts. Differing from them, our work tries to categorize Twitter users based on users' social identity or social roles. Similarly, Pirante et al. also study identity classification on Twitter \cite{priante2016whoami}. However, their approach is purely based on profile description, while we combine user self-description and tweets together. Additionally, we demonstrate that tweets are more helpful for identity classification than personal descriptions in our experiments.

In fact, learning Twitter users' identities can benefit other related tasks. Twitter is a social media where individual user accounts and organization accounts co-exist. Many user classification methods may not work on these organization accounts, e.g., gender classification. Another example is bot detection. In reality, accounts of news agencies and celebrities often look like bots \cite{chu2012detecting}, because these accounts often employ automated services or teams (so called cyborgs), and they also share features with certain classes of bots; e.g., they may be followed more than they follow. Being able to classify actors' roles on Twitter would improve our ability to automatically differentiate pure bots from celebrity accounts.


\section{Method}
\begin{figure}[!h]
    \centering
    \includegraphics[width=1\textwidth]{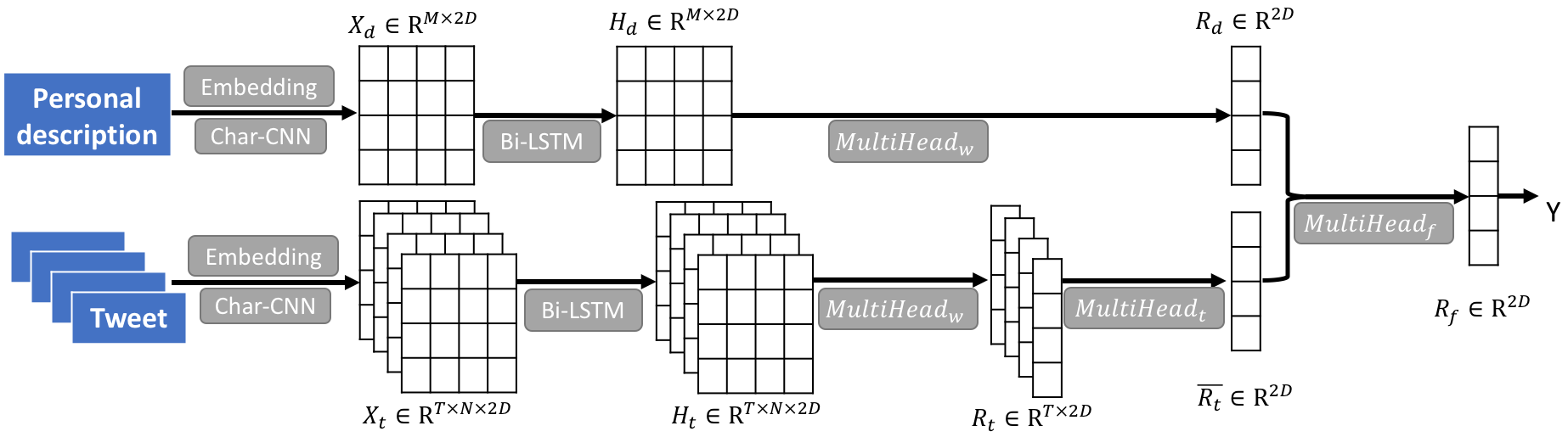}
    \caption{The architecture of hierarchical self-attention neural networks for identity classification.}
    \label{arch}
\end{figure}

In this section, we describe details of our hierarchical self-attention neural networks. The overall architecture is shown in Figure \ref{arch}. Our model first maps each word into a low dimension word embedding space, then it uses a Bidirectional Long Short-Term Memory (Bi-LSTM) network \cite{hochreiter1997long} to extract context specific semantic representations for words. Using several layers of multi-head attention neural networks, it generates a final classification feature vector. In the following parts, we elaborate these components in details. 

\subsection{Word Embedding}
Our model first maps each word in description and tweets into a word embedding space $\in R^{V\times D}$ by a table lookup operation, where $V$ is the vocabulary size, and $D$ is the embedding dimension.

Because of the noisy nature of tweet text, we further use a character-level convolutional neural network to generate character-level word embeddings, which are helpful for dealing with out of vocabulary tokens. More specifically, for each character $c_i$ in a word $w=(c_1,...,c_k)$, we first map it into a character embedding space and get $v_{c_i}\in R^{d}$. Then a convolutional neural network is applied to generate features from characters \cite{kim2014convolutional}. For a character window $v_{c_i:c_{i+h-1}}\in R^{h\times d}$, a feature $\theta_i$ is generated by $\theta_i=f(w\cdot v_{c_i:c_{i+h-1}}+b)$ where $w\in R^{h\times d}$ and $b$ are a convolution filter and a bias term respectively, $f(\cdot)$ is a non-linear function $relu$. Sliding the filter from the beginning of the character embedding matrix till the end, we get a feature vector $\boldsymbol{\theta}=[\theta_1,\theta_2,...,\theta_{k-h+1}]$. Then, we apply max pooling over this vector to get the most representative feature. With $D$ such convolutional filters, we get the character-level word embedding for word $w$.

The final vector representation $v_w\in {R^{2D}}$ for word $w$ is just the concatenation of its general word embedding vector and character-level word embedding vector. Given one description with $M$ tokens and $T$ tweets each with $N$ tokens, we get two embedding matrices $X_d \in R^{M\times 2D}$ and $X_t\in R^{T\times N\times 2D}$ for description and tweets respectively.

\subsection{Bi-LSTM}
After get the embedding matrices for tweets and description, we use a bidirectional LSTM to extract context specific features from each text. At each time step, one forward LSTM takes the current word vector $v_{w_i}$ and the previous hidden state $\overrightarrow {h_{w_{i-1}}}$ to generate the hidden state for word $w_{i}$. Another backward LSTM generates another sequence of hidden states in the reversed direction. We also tried Bi-directional GRU \cite{cho2014learning} in our initial experiments, which yields slightly worse performance.

\begin{equation}
    \begin{aligned}
    \overrightarrow {h_{w_i}}=\overrightarrow{LSTM}(v_{w_i},\overrightarrow {h_{w_{i-1}}})\\
    \overleftarrow {h_{w_i}}=\overleftarrow{LSTM}(v_{w_i},\overleftarrow {h_{w_{i+1}}})
    \end{aligned}
\end{equation}

The final hidden state $h_{w_i}\in R^{2D}$ for word $w_i$ is the concatenation of $\overrightarrow {h_{w_i}}$ and $\overleftarrow {h_{w_i}}$ as $h_{w_i} = [\overrightarrow {h_{w_i}},\overleftarrow {h_{w_i}}]$. With $T$ tweets and one description, we get two hidden state matrices $H_t\in R^{T\times N\times 2D}$ and $H_d \in R^{M\times 2D}$.

\subsection{Attention}
Following the Bi-LSTM layer, we use a word-level multi-head attention layer to find important words in a text \cite{vaswani2017attention}. 

Specifically, a multi-head attention is computed as follows:
\begin{equation*}
    \begin{aligned}
    &MultiHead(H_d) = Concat(head_1,...,head_h)W^O \\
    &head_i = softmax(\frac{H_dW_i^Q\cdot (H_dW_i^K)^T}{\sqrt{d_k}})H_dW_i^V
    \end{aligned}
\end{equation*}
where $d_k={2D}/{h}$, $W_i^Q$, $W_i^K$, $W_i^V\in R^{2D\times d_k}$, and  $W^O\in R^{hd_k\times 2D}$ are projection parameters for query, key, value, and output respectively. 

Take a user description for example. Given the hidden state matrix $H_d$ of the description, each head first projects $H_d$ into three subspaces --- query $H_dW_i^Q$, key $H_dW_i^K$, and value $H_dW_i^V$. The matrix product between key and query after softmax normalization is the self-attention, which indicates important parts in the value matrix. The multiplication of self-attention and value matrix is the output of this attention head. The final output of multi-head attention is the concatenation of $h$ such heads after projection by $W^O$.

After this word-level attention layer, we apply a row-wise average pooling to get a high-level representation vector for description. 
\begin{equation}
    R_d=row\_avg(MultiHead_w(H_d)) \in R^{2D}
\end{equation}
Similarly, we can get $T$ representation vectors from $T$ tweets using the same word-level attention, which forms $R_t\in R^{T\times 2D}$.

Further, a tweet-level multi-head attention layer computes the final tweets representation vector $\bar{R_t}$ as follows:
\begin{equation}
    \bar{R_t} = row\_avg(MultiHead_t(R_t)) \in R^{2D}
\end{equation}
In practise, we also tried using an additional Bi-LSTM layer to model the sequence of tweets, but we did not observe any significant performance gain.

Given the description representation $R_d$ and tweets representation $\bar R_t$, a field attention generates the final classification feature vector
\begin{equation}
    R_f = row\_avg(MultiHead_f([R_d; \bar R_t]))
\end{equation}
where $[R_d; \bar R_t]\in R^{2\times 2D}$ means concatenating by row.

\subsection{Final Classification}
Finally, the probability for each identity is computed by a softmax function:
\begin{equation}
    P = softmax(WR_f+b)
\end{equation}
where $W\in R^{|C|\times 2D}$ is the projection parameter, $b\in R^{|C|}$ is the bias term, and $C$ is the set of identity classes. We minimize the cross-entropy loss function to train our model,

\begin{equation}
    loss = - \sum_{c\in C}Y_{c}\cdot log P_{c}
\end{equation}
where $Y_c$ equals to 1 if the identity is of class c, otherwise 0.

\section{Experiments}
\subsection{Dataset}

To examine the effectiveness of our method, we collect two datasets from Twitter. The first is a public figure dataset. We use Twitter's verification as a proxy for public figures. These verified accounts include users in music, government, sports, business, and etc\footnote{https://help.twitter.com/en/managing-your-account/about-twitter-verified-accounts}. We sampled 156746 verified accounts and 376371 unverified accounts through Twitter's sample stream data  \footnote{https://developer.twitter.com/en/docs/tweets/sample-realtime/overview/GET\_statuse\_sample.html}. Then we collected their most recent 20 tweets from Twitter's API in November 2018. We randomly choose 5000 users as a development set and 10000 users as a test set. A summary of this dataset is shown in {Table \ref{data}}.

\begin{table}[!h]
\resizebox{\textwidth}{!}{
\begin{tabular}{|c|c|c|c|c|c|c|c|c|c|}
\hline
\multirow{2}{*}{} & \multicolumn{2}{c|}{Public Figure} & \multicolumn{7}{c|}{Identity}                                                 \\ \cline{2-10} 
                  & Verified        & Unverified       & Media & Reporter & Celebrity & Government  & Company & Sport & Regular \\ \hline
Train             & 152368          & 365749           & 1140  & 614      & 876       & 844                 & 879     & 870   & 6623   \\ \hline
Dev.              & 1452            & 3548             & 52    & 23       & 38        & 40                  & 35      & 43    & 269    \\ \hline
Test              & 2926            & 7074             & 97    & 39       & 75        & 81                  & 66      & 74    & 568    \\ \hline
\end{tabular}
}
\caption{A brief summary of our two datasets.}
\label{data}
\end{table}

In addition, we introduce another human labeled identity dataset for more fine-grained identity classification, which contains seven identity classes: ``news media'', ``news reporter'', ``government official'', ``celebrity'', ``company'', ``sport'', and ``normal people''. For each identity, we manually labelled thousands of Twitter users and collected their most recent 20 tweets for classification in November 2018. For the normal Twitter users, we randomly sampled them from the Twitter sample stream. News media accounts are these official accounts of news websites like BBC. News reporters are mainly composed of news editors or journalists. Government officials represent government offices or politicians. We collected these three types of accounts from corresponding official websites. For the other three categories, we first search Twitter for these three categories, and then we downloaded their most recent tweets using Twitter's API. Two individual workers labeled these users independently, and we include users that both two workers agreed on. The inter-rater agreement measure is 0.96. In Table \ref{handle}, we list several representative Twitter handles for each identity class except for normal users. Table \ref{data} shows a summary of this dataset. We randomly select 500 and 1000 users for development and test respectively. Since normal users are the majority of Twitter users, about half of the users in this dataset are normal users.

\begin{table}[!h]
\resizebox{\textwidth}{!}{
\begin{tabular}{|l|l|l|l|l|l|}
\hline
News Media     & News Reporter &Celebrity    & Government Official & Company      & Sport    \\ \hline
CBSNews       & PamelaPaulNYT &aliciakeys   & USDOL               & VisualStudio & NBA      \\ \hline
earthtimes     & HowardKurtz &Adele        & RepRichmond       & lifeatgoogle & Pirates  \\ \hline
BBCNewsAsia    & jennaportnoy &GreenDay     & HouseGOP            & BMW          & NFL      \\ \hline
phillydotcom   & wpjenna    &ladygaga     & BelgiumNATO         & AEO          & KKRiders \\ \hline
TheSiasatDaily & twithersAP &TheEllenShow & usafpressdesk       & Sony         & USAGym   \\ \hline
\end{tabular}
}
\caption{Five representative Twitter handles for each identity class except for regular users.}
\label{handle}
\end{table}

This paper focuses on a content-based approach for identity classification, so we only use personal description and text of each tweet for each user.

\subsection{Hyperparameter Setting}
In our experiments, we initialize the general word embeddings with released 300-dimensional Glove vectors\footnote{https://nlp.stanford.edu/projects/glove/} \cite{pennington2014glove}. For words not appearing in Glove vocabulary, we randomly initialize them from a uniform distribution $U(-0.25,0.25)$. The 100-dimensional character embeddings are initialized with a uniform distribution $U(-1.0,1.0)$. These embeddings are adapted during training. We use filter windows of size 3,4,5 with 100 feature maps each.  The state dimension $D$ of LSTM is chosen as 300. For all the multi-head attention layers, we choose the number of heads as 6. We apply dropout \cite{srivastava2014dropout} on the input of Bi-LSTM layer and also the output of the softmax function in these attention layers. The dropout rate is chosen as 0.5. The batch size is 32. We use Adam update rule \cite{kingma2014adam} to optimize our model. The initial learning rate is $10^{-4}$ and it drops to $10^{-5}$ at the last 1/3 epochs. We train our model 10 epochs, and every 100 steps we evaluate our method on development set and save the model with the best result. All these hyperparameters are tuned on the development set of identity dataset. We implemented our model using Tensorflow \cite{abadi2016tensorflow} on a Linux machine with Titan XP GPUs.

\subsection{Baselines}
MNB: Multinomial Naive Bayes classifier with unigrams and bigrams. The term features are weighted by their TF-IDF scores. Additive smoothing parameter is set as $10^{-4}$ via a grid search on the development set of identity dataset.\\
SVM: Support Vector Machine classifier with unigrams and linear kernel. The term features are weighted by their TF-IDF scores. Penalty parameter is set as 100 via a grid search on the development set of identity dataset.\\
CNN: Convolutional Neural Networks \cite{kim2014convolutional} with filter window size 3,4,5 and 100 feature maps each. Initial learning rate is $10^{-3}$ and drops to $10^{-4}$ at the last 1/3 epochs.\\
Bi-LSTM: Bidirectional-LSTM model with 300 hidden states in each direction. The average of output at each step is used for the final classification. \\
Bi-LSTM-ATT: Bidirectional-LSTM model enhanced with self-attention. We use multi-head attention with 6 heads. \\
fastText \cite{joulin2016bag}: we set word embedding size as 300, use unigram, and train it 10 epochs with initial learning 1.0.\\
For methods above, we combine personal description and tweets into a whole document for each user.

\subsection{Results}

In Table \ref{result}, we show comparison results between our model and baselines. Generally, LSTM based methods work the best among all these baseline approaches. SVM has comparable performance to these neural network based methods on the identity dataset, but falls behind on the larger public figure dataset.

\begin{table*}[!ht]
\centering
\begin{tabular}{|l|l|c|c|c|c|}
\hline
\multicolumn{2}{|l|}{\multirow{2}{*}{}}               & \multicolumn{2}{c|}{Public Figure}                                & \multicolumn{2}{c|}{Identity}                                     \\ \cline{3-6} 
\multicolumn{2}{|l|}{}                                & Accuracy                        & Macro-F1                              & Accuracy                       & Macro-F1                               \\ \hline
Baselines                       & MNB                 & 81.81                           & 82.79                           & 82.9                           & 75.91                            \\ \cline{2-6} 
                                & SVM                 & 90.60                           & 88.59                           & 85.9                           & 80.19                            \\ \cline{2-6} 
                                & fastText            & 90.93                           & 89.01                           & 85.7                           & 80.01                            \\ \cline{2-6} 
                                & CNN                 & 91.45                           & 89.85                           & 85.9                           & 81.24                            \\ \cline{2-6} 
                                & Bi-LSTM             & 93.10                           & 91.84                           & 86.5                           & 84.25                            \\ \cline{2-6} 
                                & Bi-LSTM-ATT         & 93.23                           & 91.94                           & 87.3                           & 83.35                            \\ \hline
\multirow{4}{*}{Ablated Models} & w/o attentions      & 93.78                           & 92.45                           & 87.0                           & 83.26                            \\ \cline{2-6} 
                                & w/o charcnn         & 93.47                           & 92.23                           & 89.0                           & 85.39                            \\ \cline{2-6} 
                                & w/o description     & 92.39                           & 90.90                           & 86.7                           & 81.56                            \\ \cline{2-6} 
                                & w/o tweets          & 91.62                           & 89.77                           & 84.2                           & 78.41                            \\ \hline
                                & Full Model         & \textbf{94.21} & \textbf{93.07} & \textbf{89.5} & \textbf{86.09} \\ \hline
                                 & Full Model-transfer      &  & & \textbf{91.6} & \textbf{88.63} \\ \hline
\end{tabular}
\caption{Comparisons between our methods and baselines.}
\label{result}
\end{table*}

Our method outperforms these baselines on both datasets, especially for the more challenging fine-grained identity classification task. Our model can successfully identify public figures with accuracy 94.21\% and classify identity with accuracy 89.5\%. Compared to a strong baseline Bi-LSTM-ATT, our model achieves a 2.2\% increase in accuracy, which shows that our model with structured input has better classification capability.

We further performed ablation studies to analyze the contribution of each model component, where we removed attention modules, character-level word embeddings, tweet texts, and user description one by one at a time. As shown in Table \ref{result}, attention modules make a great contribution to the final classification performance, especially for the more fine-grained task. We present the performance breakdown for each attention module in Table \ref{att}. Each level of attention effectively improves the performance of our model. Recognizing important words, tweets, and feature fields at different levels is helpful for learning classification representations. According to Table \ref{result}, the character-level convolutional layer is also helpful for capturing some character-level patterns. 

We also examined the impact of two different text fields: personal description and tweets. Indeed, we found that what users tweeted about is more important than what they described themselves. On both datasets, users' tweets provide more discriminative power than users' personal descriptions.

\begin{table}[!h]
\centering
\begin{tabular}{|l|c|c|}
\hline
               & Accuracy & Macro-F1 \\ \hline
Full Model     & 89.5     & 86.09    \\ \hline
w/o word attention  & 88.8     & 84.41    \\ \hline
w/o field attention & 88.5     & 85.24    \\ \hline
w/o tweet attention & 88.5     & 84.6     \\ \hline
w/o all attention   & 87.0     & 83.26    \\ \hline
\end{tabular}
\caption{The effectiveness of different levels of attentions tested on the identity dataset.}
\label{att}
\end{table}

\subsection{Transfer Learning for Fine-grained Identity Classification}

In reality, it is expensive to get a large-scale human labeled dataset for training a fine-grained identity classifier. However, a well-known drawback of neural network based methods is that they require a lot of data for training. Recently, learning from massive data and transferring learned knowledge to other tasks attracts a lot of attention \cite{peters2018deep,devlin2018bert}. Since it is relatively easier to get a coarse-grained identity dataset to classify those public figures, we explore how to use this coarse-grained public figure dataset to help the training of fine-grained identity classifier.

\begin{figure*}[!h]
    \centering
    \includegraphics[width=0.8\textwidth]{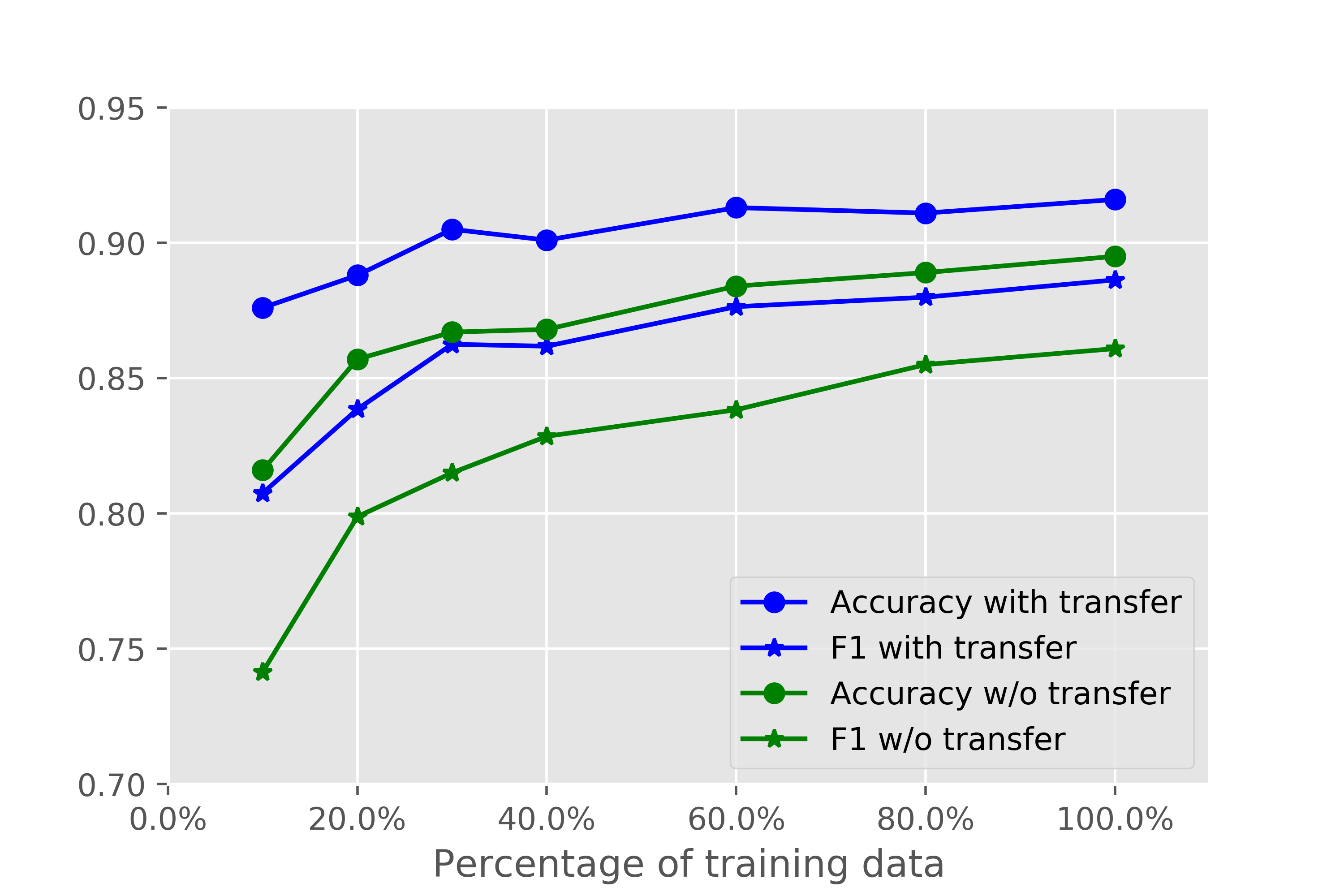}
    \caption{Performance comparison between our model with transfer learning and without. We train our model on various amounts of training data. }
    \label{pretrain}
\end{figure*}

Specifically, we first pretrain a binary classifier on the public figure dataset and save the best trained model on its development set. To make a fair comparison, we excluded all the users appearing in identity dataset from the public figure dataset when we built our datasets. Then we initialize the parameters of fine-grained identity classifier with this pretrained model except for the final classification layer. After such initialization step, we first train the final classification layer for 3 epochs with learning rate 0.01, and then train our full identity classification model with the same procedure as before. We observe a big performance boost when we apply such pretraining as shown in Table \ref{result}. The classification accuracy for the fine-grained task increases by 2.1\% with transfer learning.

We further examined the performance of our model with pretraining using various amounts of training data. As shown in Figure \ref{pretrain}, our pretrained model reaches a comparable performance only with 20\%-30\% labeled training data when compared to the model trained on full identity dataset without pretraining. Using only 20\% of training data, we can get accuracy 0.888 and F1 0.839. If we increase the data size to 30\% of the training data, the accuracy and F1 will increase to 0.905 and 0.863 respectively. Such pretraining makes great improvements over fine-grained identity classification especially when we lack labeled training data.

\subsection{Case Study}

In this section, we present a case study in the test set of identity dataset to show the effectiveness of our model. Because of the difficulties of visualizing and interpreting multi-head attention weights, we instead average over the attention weights in multiple heads which gives us an approximation of the importance of each word in texts. Take the user description for example, the approximated importance weight of each word in the description is given by
\begin{align*}
    \alpha_d = row\_avg(\frac{1}{h} \sum_i softmax(\frac{H_dW_i^Q (H_dW_i^K)^T}{\sqrt{d_k}})) 
\end{align*}
Similarly, we can get the importance weights for tweets as well as words in tweets.

\begin{figure*}[!h]
    \centering
    \includegraphics[width=1.0\textwidth]{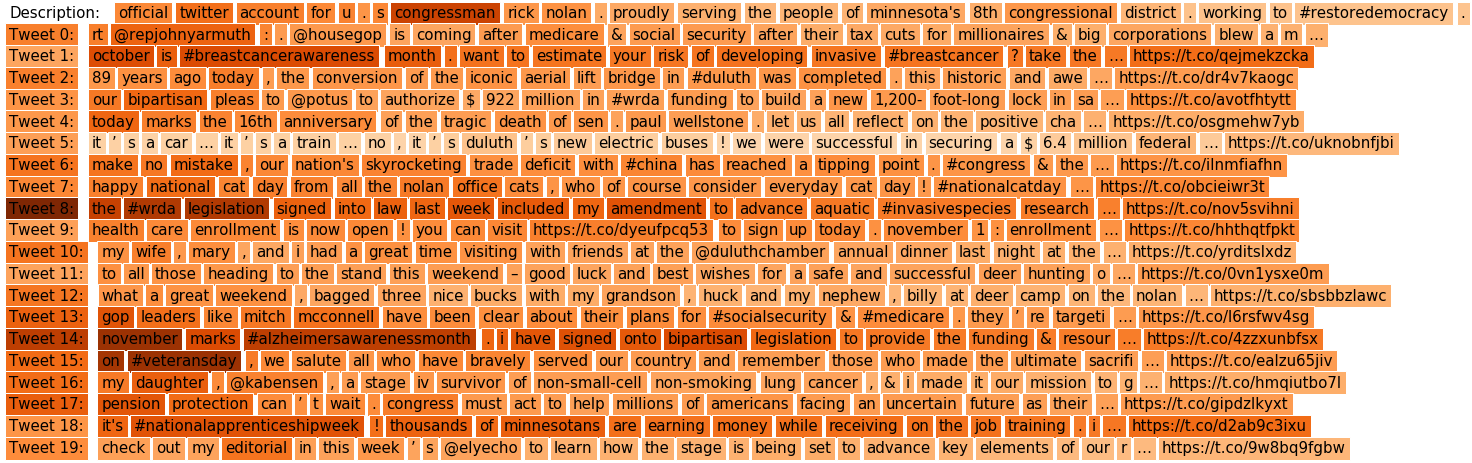}
    \caption{The visualization of attention weights for each tweet and description. The color depth denotes the importance degree of a word per tweet. The importance of each tweet is depicted as the background color of corresponding tweet header. }
    \label{case}
\end{figure*}

In Figure \ref{case}, we show twenty tweets and a description from a government official user.
We use the background color to represent importance weight for each word. The color depth denotes the importance degree of a word per tweet. We plot the tweet-level importance weights as the background color of tweet index at the beginning of each tweet. As shown in this figure, words like ``congressman'', ``legislation'' in this user's description are important clues indicating his/her identity. From the tweet-level attention, we know that 8th and 14th tweets are the most important tweets related with the identity because they include words like ``legislation'' and ``bipartisan''. On the contrary, 5th tweet of this user only contain some general words like ``car'', which makes it less important than other tweets.

\subsection{Error Analysis}
We perform an error analysis to investigate why our model fails in certain cases. Table \ref{confusion} shows the confusion matrix generated from prediction results of our identity dataset. As shown in this table, it is relatively harder for our model to distinguish between celebrities and regular users. We further looked at such errors with high confidences and found that some celebrities just have not posted any indicating words in their tweets or descriptions. For example, one celebrity account only use ``A Virgo'' in the description without any other words, which makes this account predicted as a regular user. Including other features like number of followers or network connections may overcome this issue, and we leave it for future work. Another common error happens when dealing with non-English tweets. Even enhanced with transferred knowledge from the large-scale verify dataset, our model still cannot handle some rare languages in the data.

\begin{table}[]
\centering
\begin{tabular}{l|l|l|l|l|l|l|l}
\hline
   \backslashbox{Truth}{Prediction}       & Regular & Media & Celebrity & Sport & Company & Government & Reporter \\ \hline
Regular    & 535    & 10    & 12        & 0     & 5       & 2    & 4        \\ \hline
Media     & 6      & 81    & 1         & 4     & 2       & 2    & 1        \\ \hline
Celebrity & 15     & 0     & 55        & 2     & 2       & 1    & 0        \\ \hline
Sport     & 1      & 1     & 0         & 71    & 1       & 0    & 0        \\ \hline
Company   & 1      & 2     & 1         & 4     & 58      & 0    & 0        \\ \hline
Government      & 1      & 1     & 0         & 0     & 0       & 79   & 0        \\ \hline
Reporter  & 1      & 0     & 1         & 0     & 0       & 0    & 37       \\ \hline
\end{tabular}
\caption{The confusion matrix generated by our best trained model from the test set of identity dataset}
\label{confusion}
\end{table}

\section{Discussion \& Conclusion}
As previously discussed, identities can vary in granularity. We examined two levels - coarse grained (verified or not) and more fine grained (news media, government officials, etc.). However, there could be more levels.  This limits our understanding of activities of online actors with those identities.  A hierarchical approach for identity classification might be worth further research.  Future research should take this into consideration and learn users' identities in a more flexible way. Besides, because of the nature of social media, the content on Twitter would evolve rapidly. In order to deploy our method in real-time, we need consider an online learning procedure that adapts our model to new data patterns. Since our method is purely content-based, potential improvements could be made using additional information like the number of users' followers, users' network connections, and even their profile images. We leave this as our future work.  

In the real-world people often have multiple identities - e.g., Serbian, Entrepreneur, Policewoman, Woman, Mother. The question is what is the relation between identities, users, and user accounts.  Herein, we treat each account as a different user.  However, in social media, some people use different accounts and/or different social media platforms for different identities - e.g., Facebook for Mother, Twitter for Entrepreneur and a separate Twitter handle for official policewoman account. In this paper, we made no effort to determine whether an individual had multiple accounts.  Thus, the same user may get multiple classifications if that user has multiple accounts. Future work should explore how to link multiple identities to the same user. To this point, when there is either a hierarchy of identities or orthogonal identity categories, then using identities at different levels of granularity, as we did herein, enables multiple identities to be assigned to the same account and so to the same user.

In conclusion, we introduce two datasets for online user identity classification. One is automatically extracted from Twitter, the other is a manually labelled dataset. We present a novel content-based method for classifying social media users into a set of identities (social roles) on Twitter. Our experiments on two datasets show that our model significantly outperforms multiple baseline approaches. Using one personal description and up to twenty tweets for each user, we can identify public figures with accuracy 94.21\% and classify more fine-grained identities with accuracy 89.5\%. We proposed and tested a transfer learning scheme that further boosts the final identity classification accuracy by a large margin. Though, the focus of this paper is learning users' social identities. It is possible to extend this work to predict other demographics like gender and age.

\nocite{*}
\bibliography{reference}
\bibliographystyle{splncs04}

\end{document}